\documentclass[referee]{aa} 

\usepackage{graphicx}
\usepackage{txfonts}
\bibpunct{(}{)}{;}{a}{}{,} 

\begin{document} 

   \title{Revisiting TrES-5 b: departure from a linear ephemeris instead of short-period transit timing variation\thanks{This research is partly based on (1) data obtained at the 1.5m telescope of the Sierra Nevada Observatory (Spain), which is operated by the Consejo Superior de Investigaciones Cient\'{\i}ficas (CSIC) through the Instituto de Astrof\'{\i}sica de Andaluc\'{\i}a, and (2) observations collected with telescopes at the Rozhen National Astronomical Observatory.}\thanks{The light curves are available at the CDS.}}

   \subtitle{}

   \author{G.~Maciejewski\inst{1}
          \and
          M.~Fern\'andez\inst{2}
          \and
          F.~Aceituno\inst{2}
          \and
          J.~L.~Ramos\inst{2}
          \and
          D.~Dimitrov\inst{3}
          \and
          Z.~Donchev\inst{3}
          \and
          J.~Ohlert\inst{4,5}
          }

   \institute{Institute of Astronomy, Faculty of Physics, Astronomy and Informatics,
              Nicolaus Copernicus University in Toru\'n, Grudziadzka 5, 87-100 Toru\'n, Poland\\
              \email{gmac@umk.pl}
         \and
             Instituto de Astrof\'isica de Andaluc\'ia (IAA-CSIC), 
             Glorieta de la Astronom\'ia 3, 18008 Granada, Spain
         \and
             Institute of Astronomy, Bulgarian Academy of Sciences, 
             72 Tsarigradsko Chausse Blvd., 1784 Sofia, Bulgaria
         \and
             Michael Adrian Observatorium, Astronomie Stiftung Trebur, 
             65428 Trebur, Germany
         \and
             University of Applied Sciences, Technische Hochschule Mittelhessen, 
             61169 Friedberg, Germany
             }
   \authorrunning{G.~Maciejewski et al.}
   \date{Received --- 2021; accepted --- 2021}
 
 
  \abstract
   {}
   {The orbital motion of the transiting hot Jupiter TrES-5~b was reported to be perturbed by a planetary companion on a nearby orbit. Such compact systems do not frequently occur in nature, and learning their orbital architecture could shed some light on hot Jupiters' formation processes.}
   {We acquired fifteen new precise photometric time series for twelve transits of TrES-5~b between June 2019 and October 2020 using 0.9-2.0 m telescopes. The method of precise transit timing was employed to verify the deviation of the planet from the Keplerian motion.}
   {Although our results show no detectable short-time variation in the orbital period of TrES-5~b and the existence of the additional nearby planet is not confirmed, the new transits were observed about two minutes earlier than expected. We conclude that the orbital period of the planet could vary in a long timescale. We found that the most likely explanation of the observations is the line-of-sight acceleration of the system's barycentre due to the orbital motion induced by a massive, wide-orbiting companion.}
   {}


   \maketitle
%

\section{Introduction}\label{Sect:Intro}

The star GSC 3949-967 is a cool dwarf that hosts a transiting hot Jupiter on a 1.48~d orbit \citep{2011ApJ...741..114M}. Its effective temperature $T_{\rm eff} = 5171 \pm 36$ K corresponds to the K1 spectral type \citep{2013ApJS..208....9P}\footnote{updated
values at https://www.pas.rochester.edu/$\sim$emamajek/EEM\_dwarf\_UBVIJHK\_colors\_Teff.txt}. The planet was discovered as the fifth planet in Trans-atlantic Exoplanet Survey \citep[TrES,][]{2007ASPC..366...13A}. It was found to have a mass $M_{\rm b} \approx 1.8$ $M_{\rm Jup}$ and a radius $R_{\rm b} \approx 1.2$ $R_{\rm Jup}$ \citep{2011ApJ...741..114M,2015MNRAS.448.2617M,2016AcA....66...55M}. Using 1.2--2.5 m telescopes, \citet{2015MNRAS.448.2617M} acquired precise photometric time series for six transits, the mid-points of which were found to be consistent with a linear ephemeris. The same conclusion was reached by \citet{2016AcA....66...55M} as a result of a homogenous analysis of photometric data enhanced with four transit light curves from 0.6--2.0 m instruments. \citet{2018MNRAS.480..291S} obtained additional 30 transit light curves, mostly acquired with 0.3--0.8 m telescopes, and found variations in transit times modulated with an amplitude of $\approx 2$ minutes and periodicity of $\approx 100$ days. Numerical simulations showed that a Saturn-mass planet on an outer orbit could induce such a signal if both planets were close to the 2:1 orbital period commensurability.

Statistical studies show that hot Jupiters are usually devoid of planetary companions on nearby orbits \citep{2009ApJ...693.1084W,2011ApJ...732L..24L,2012PNAS..109.7982S} and can be accompanied by massive planets on wide and eccentric orbits \citep[e.g.][]{2017A&A...602A.107B}. This picture aligns with planet formations theories that invoke a high-eccentricity migration as a mechanism bringing hot Jupiters from their birthplaces beyond a water frost line to the tight orbits observed nowadays. Being perturbed by the additional nearby companion planet, TrES-5~b would be the fourth known hot Jupiter in a compact planetary system, similar to WASP-47 \citep{2015ApJ...812L..18B}, Kepler-730 \citep{2019ApJ...870L..17C}, and TOI-1130 \citep{2020ApJ...892L...7H}.   

\section{Observations and data reduction}\label{Sect:Obs}

Fifteen new precise photometric time series for twelve transits of TrES-5~b were acquired between June 2019 and October 2020 by engaging four instruments:
\begin{itemize}
	\item the 1.5 m Ritchey-Chr\'etien telescope (OSN150) at the Sierra Nevada Observatory (OSN, Spain) equipped with a Roper Scientific VersArray 2048B CCD camera,
	\item the 2.0 m Ritchey-Chr\'etien-Coud\'e telescope (ROZ) at the National Astronomical Observatory Rozhen (Bulgaria) with a Roper Scientific VersArray 1300B CCD camera,
	\item the 1.2 m Trebur one-meter telescope (TRE) at the Michael Adrian Observatory in Trebur (Germany) with an SBIG STL-6303 CCD camera, and
	\item the 0.9 m Ritchey-Chr\'etien telescope (OSN90) at OSN with a Roper Scientific VersArray 2048B CCD camera.
\end{itemize}
Observing runs were scheduled with time margins of 45-90 minutes before the beginning and after the end of the expected transit in order to probe photometric trends. However, portions of this out-of-transit monitoring were lost in some cases due to unfavourable weather conditions or scheduling constraints. The instruments were automatically or manually guided to keep the star at the same position in the CCD matrix. They were also mildly defocused to allow for longer exposure times which minimised the observing time lost for CCD readout. Most of the light curves were acquired in white light, i.e.\ without any filter, to increase the signal-to-noise ratio for transit timing purposes. Details on the individual runs are given in Table~\ref{tab.Obs}.

\begin{table*}[h]
\caption{Details on the observing runs.} 
\label{tab.Obs}      
\centering                  
\begin{tabular}{l l c c c c c c c}      
\hline\hline                
Date UT (Epoch)  & Telescope & Band & UT start-end  &  $X$                                & $N_{\rm{obs}}$ & $t_{\rm{exp}}$ (s) & $\Gamma$ & pnr (ppth)\\
\hline
2019 Jul 09 (2180) & OSN150 & clear & 22:50--03:27 & $1.20 \rightarrow 1.08 \rightarrow 1.15$ & 667 & 20 & 2.69 & 0.76	\\
2019 Aug 24	(2211) & OSN150 & clear & 21:45--00:07 & $1.09 \rightarrow 1.08 \rightarrow 1.34$ & 703 & 20 & 2.69 & 0.71	\\
2019 Oct 18	(2248) & OSN150 & clear & 18:38--22:06 & $1.08 \rightarrow 1.28$                  & 288 & 40 & 1.42 & 0.82	\\
                   & OSN90 & clear & 18:50--22:06 & $1.08 \rightarrow 1.28$                  & 253 & 40 & 1.31 & 1.49	\\
2019 Nov 30	(2277) & OSN150 & clear & 18:52--21:56 & $1.23 \rightarrow 1.91$                  & 496 & 20 & 2.69 & 0.83	\\
2020 Jul 15 (2431) & OSN150 & clear & 23:59--04:01 & $1.10 \rightarrow 1.08 \rightarrow 1.25$ & 650 & 20 & 2.69 & 0.64	\\
2020 Jul 24 (2437) & ROZ & $R$   & 21:46--01:19 & $1.06 \rightarrow 1.05 \rightarrow 1.17$ & 493 & 20 & 2.40 & 0.73	\\
                   & OSN150 & clear & 21:30--01:40 & $1.23 \rightarrow 1.08 \rightarrow 1.11$ & 613 & 20 & 2.69 & 0.77	\\
2020 Jul 27 (2439) & ROZ & $R$   & 21:11--23:37 & $1.07 \rightarrow 1.05 \rightarrow 1.08$ & 321 & 20 & 2.39 & 0.87	\\
                   & TRE & clear & 21:01--01:00 & $1.08 \rightarrow 1.01 \rightarrow 1.04$ & 108 & 80 & 0.67 & 1.27	\\
2020 Sep 02 (2464) & OSN150 & clear & 21:53--02:46 & $1.08 \rightarrow 1.61$                  & 784 & 20 & 2.69 & 0.80	\\
2020 Sep 05 (2466) & OSN150 & clear & 19:43--01:18 & $1.14 \rightarrow 1.08 \rightarrow 1.34$ & 816 & 20 & 2.37 & 0.79	\\
2020 Sep 08 (2468) & TRE & clear & 20:01--00:31 & $1.02 \rightarrow 1.01 \rightarrow 1.21$ & 182 & 80 & 0.67 & 1.31	\\
2020 Oct 24 (2499) & OSN150 & clear & 18:55--23:06 & $1.09 \rightarrow 1.54$                  & 546 & 25 & 2.19 & 0.95	\\
2020 Oct 27 (2501) & OSN150 & clear & 18:43--22:02 & $1.09 \rightarrow 1.37$                  & 335 & 30 & 1.69 & 0.80	\\\hline                                   
\end{tabular}
\tablefoot{Date UT is given for the beginning of an observing run. Epoch is the transit number from the initial ephemeris given in \citet{2011ApJ...741..114M}. $X$ tracks the target's airmass during a run. $N_{\rm{obs}}$ is the number of useful scientific exposures. $t_{\rm{exp}}$ is exposure time. $\Gamma$ is the median number of exposures per minute. $pnr$ is the photometric noise rate \citep{2011AJ....142...84F} in parts per thousand (ppth) of the normalised flux per minute of observation.}
\end{table*}

The science frames were the subject of a standard calibration and reduction procedure which is implemented in AstroImageJ software \citep{2017AJ....153...77C}. The OSN and Rozhen observations were de-biased. In the case of the Trebur data, a dark-current correction was applied. Flat fielding was performed using sky (mainly) or dome (occasionally) flat frames. A built-in procedure was used to convert timestamps into barycentric Julian dates and barycentric dynamical time $\rm{BJD_{TDB}}$.

\begin{figure}
	\includegraphics[width=0.5\columnwidth]{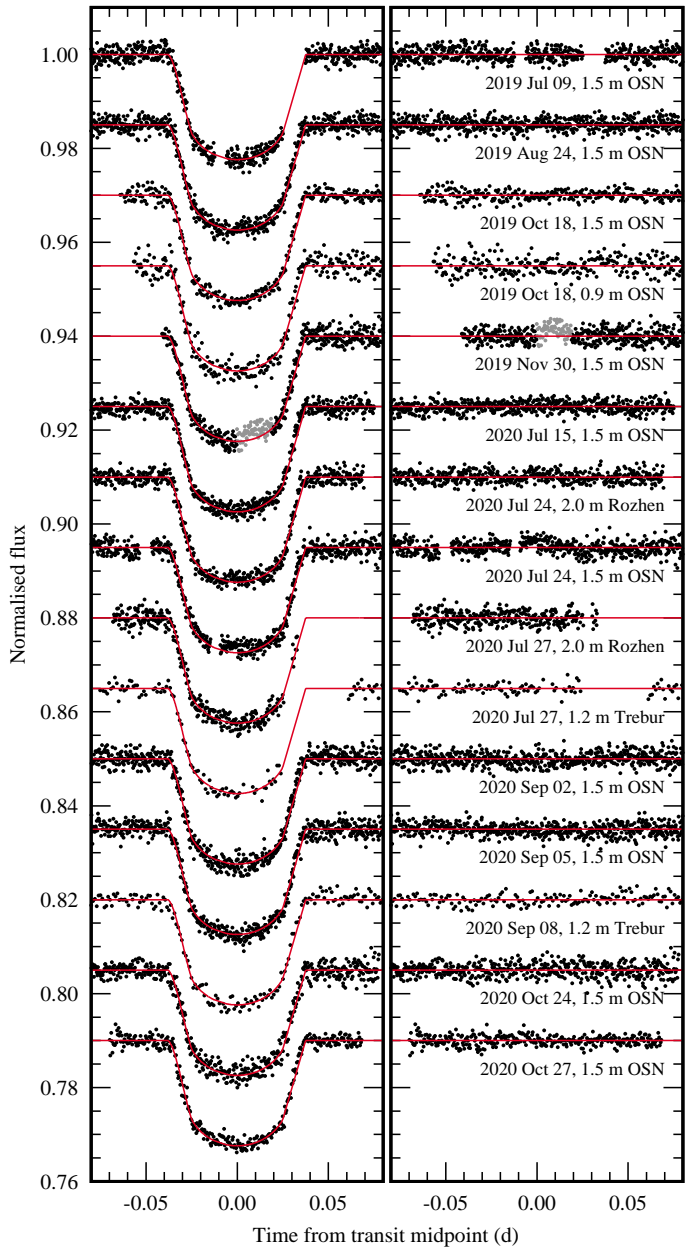}
    \caption{Left: new transit light curves for TrES-5~b, sorted by the observation date and, if observed simultaneously, by the telescope's aperture size. The best-fitting model is drawn with red lines. A signature of star-spot occultation identified in the light curve acquired on 2019 November 30 is marked with grey points. These measurements were masked out in transit modelling. Right: photometric residuals from the transit model.}
    \label{fig:lcs}
\end{figure}

Fluxes were obtained with the differential aperture photometry method. Aperture size and a collection of comparison stars were optimised to achieve the lowest data point scatter in the light curves. The photometric time series were tested for trends against airmass, time, and seeing along with a trial transit model. Then fluxes were normalised to unity outside the transit. The individual transit light curves are shown in Fig.~\ref{fig:lcs}. 

\section{Light curve analysis}\label{Sect:LCanalysis}

Transit light curves were analysed with the Transit Analysis Package \citep[TAP,][]{2012AdAst2012E..30G} which uses the analytical approach of \citet{2002ApJ...580L.171M} to model a transit signature and a wavelet-based likelihood function \citep{2009ApJ...704...51C} to estimate the parameter uncertainties. A transit model was parametrised with an orbital inclination $i_{\rm{b}}$, a semi-major axis scaled in stellar radii $a_{\rm{b}}/R_{\star}$, a ratio of planet to star radii $R_{\rm{b}}/R_{\star}$, limb darkening (LD) coefficients of a quadratic law, and a time of transit midpoint $T_{\rm{mid}}$. The first three parameters were linked for all the light curves. The LD coefficients were linked for light curves acquired in the same passbands, and the values of $T_{\rm{mid}}$ were linked for the same epochs. A second-order polynomial was allowed to be fitted separately for each light curve along with the joint transit model to account for possible trends in the time domain and their uncertainties. 

In a trial iteration, the values of $R_{\rm{b}}/R_{\star}$ were allowed to vary independently for each epoch to account for possible variations due to starspots. All determinations of $R_{\rm{b}}/R_{\star}$ were found to be consistent with a weighted mean value of up to $2.5\sigma$ leaving no space for detection of any transit depth variation.

The values of LD coefficients were interpolated from the tables of \citet{2011A&A...529A..75C} for stellar parameters reported by \citet{2011ApJ...741..114M}. Their values for white light data were calculated as averages of their values in B, V, R, and I filters which cover the spectral sensitivity of the instrumental setup. In a trial iteration, the values of the LD coefficients were allowed to be free parameters of a fit. Their values were found to agree well within the $1\sigma$ range with the theoretical predictions. However, their uncertainties between 0.1 and 0.3 were considered insufficiently precise. In the final iteration, the LD coefficients were allowed to vary around their theoretical values under a Gaussian penalty with a conservative value of 0.1. 

The coefficients of the second-order polynomials, which accounted for possible trends in the time domain, were consistent with zero primarily within the 1-2$\sigma$ range, rarely reaching 2-3$\sigma$.

The refined transit parameters were determined as medians from the posterior parameter distributions generated by ten random walk chains, each $10^6$ steps long with a 10\% burn-in phase. The 15.9 and 84.1 percentiles of the marginalised posterior probability distributions were used to determine $1\sigma$ uncertainties of the parameters.

\section{Results}\label{Sect:Results}

\subsection{System parameters}\label{SSect:Results.Parameters}

The best-fitting transit model is plotted for individual light curves in Fig.~\ref{fig:lcs}. The refined transit parameters are listed in Table~\ref{tab.Pars} together with additional quantities which can be calculated using standard formulae directly from the transit model. There are also the literature values that are shown for comparison purposes. 

\begin{table*}[h]
\caption{Refined and literature system parameters from new TrES-5 transits.} 
\label{tab.Pars}      
\centering                  
\begin{tabular}{l c c c c c}      
\hline\hline                
 Parameter & This paper & (1) & (2) & (3) & (4)\\
\hline
Orbital inclination, $i_{\rm{b}}$ $(^{\circ})$ & $84.69^{+0.15}_{-0.14}$ & $84.529\pm0.005$ & $84.27 \pm 0.26$ & $84.65^{+0.24}_{-0.22}$  & $85.78 \pm 0.39$\\
Scaled semi-major axis, $a_{\rm{b}}/R_{\star}$ & $6.190^{+0.050}_{-0.049}$ & $6.074\pm0.143$ & $6.10 \pm 0.11$ & $6.188^{+0.085}_{-0.078}$  & --\\
Radii ratio, $R_{\rm{b}}/R_{\star}$ & $0.14142^{+0.00063}_{-0.00066}$ & $0.1436\pm0.0012$ & $0.143 \pm 0.0012$ & $0.14203^{+0.00084}_{-0.00091}$  & $0.1405 \pm 0.0011$\\
Transit depth, $\delta$ (ppth) & $20.00^{+0.18}_{-0.19}$ & $20.62\pm0.35$ & $20.45\pm0.35$ & $20.17^{+0.24}_{-0.26}$  & $19.74 \pm 0.31$\\
Impact parameter, $b$ $(R_{\star})$ & $0.573^{+0.017}_{-0.016}$ & $0.579\pm0.026$ & -- & $0.577^{+0.025}_{-0.027}$  & --\\
Transit duration, $T_{14}$ $(min)$ & $108.4^{+1.4}_{-1.3}$ & -- & -- & --  & --\\
Stellar density, $\rho_{\star}$ $(\rho_{\odot})$ & $1.450^{+0.035}_{-0.034}$ & -- & $1.381 \pm 0.051$ & $1.449^{+0.060}_{-0.055}$  & --\\
\hline
\end{tabular}
\tablefoot{References: (1) -- \citet{2011ApJ...741..114M}, (2) -- \citet{2015MNRAS.448.2617M}, (3) -- \citet{2016AcA....66...55M}, (4) -- \citet{2018MNRAS.480..291S}.}
\end{table*}

For almost all parameters, our determinations agree well within the $1\sigma$ range with the values reported in the previous studies. They are also the most precise results obtained so far. The only exception is the orbital inclination reported in \citet{2011ApJ...741..114M} the error of which is likely underestimated. 

\subsection{Transit timing}\label{SSect:Results.Timing}

\begin{figure}
	\includegraphics[width=0.5\columnwidth]{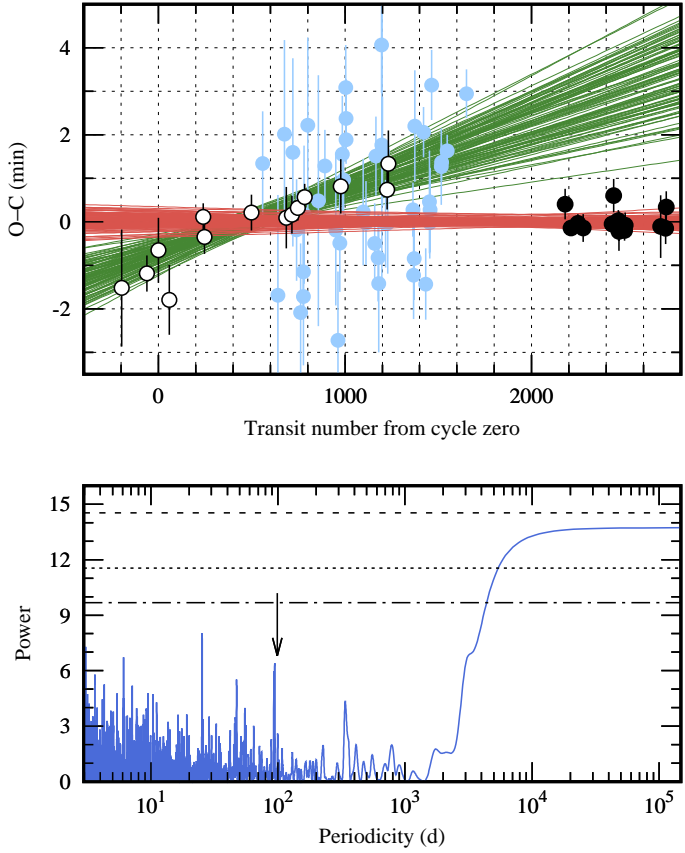}
    \caption{Upper panel: transit timing residuals against the refined linear ephemeris. The new observations are marked with filled black dots, and the mid-transit times from \citet{2016AcA....66...55M} are marked with open circles. The light blue filled dots show data from \citet{2018MNRAS.480..291S}. The uncertainties of the refined ephemeris are illustrated with a bunch of 100 red lines drawn from the Markov chains. The green lines show the ephemeris from \citet{2016AcA....66...55M} which is brought up here for comparison purposes. Lower panel: AoV periodogram generated for transit timing residuals. The dashed horizontal lines place the empirical FAP levels of 5\%, 1\%, and 0.1\% (from the bottom up). The location of the periodicity of $\sim$99 days that was claimed by \citet{2018MNRAS.480..291S} is marked with an arrow.}
    \label{fig:ttres}
\end{figure}

The new mid-transit times, which are listed in Table~\ref{tab.TTimes}, were combined with literature data compiled by \citet{2016AcA....66...55M} to refine a transit ephemeris in the form
\begin{equation}
  T_{\rm{mid}}= T_0 + P_{\rm{b}} E,
\end{equation}
where $T_{\rm{mid}}$ is the time of the $E$-th transit counted from the cycle-zero epoch $T_0$ given in \citet{2011ApJ...741..114M}, and $P_{\rm{orb}}$ is the orbital period. The data from \citet{2018MNRAS.480..291S} were skipped in the final analysis because they lag behind the other datasets in terms of timing accuracy. The best-fitting values of $T_0$ and $P_{\rm{orb}}$ and their uncertainties were inferred from the posterior probability distributions of those parameters generated with the MCMC algorithm, running 100 chains, each of which was $10^4$ steps long after discarding the first 1000 trials. We obtained $T_0 = 2455443.25321 \pm 0.00011$ $\rm{BJD_{TDB}}$ and $P_{\rm{orb}} = 1.482246502 \pm 0.000000051$ d. The model yields $\chi^2 = 32.9$ with 24 degrees of freedom. The timing residuals against the refined transit ephemeris are plotted in the upper panel of Fig.~\ref{fig:ttres}.

\begin{table}[h]
\caption{Mid-transit times for the new transit light curves.} 
\label{tab.TTimes}      
\centering                  
\begin{tabular}{r c c c c}      
\hline\hline                
Epoch  & $T_{\rm{mid}}$ $(\rm{BJD_{TDB}})$ & $+\sigma$ (d) & $-\sigma$ (d)  &  $N_{\rm {lc}}$\\
\hline
2180 & 2458674.55087 & 0.00024 & 0.00024 & 1 \\
2211 & 2458720.50013 & 0.00012 & 0.00012 & 1 \\
2248 & 2458775.34333 & 0.00014 & 0.00013 & 2 \\
2277 & 2458818.32840 & 0.00022 & 0.00023 & 1 \\
2431 & 2459046.59442 & 0.00014 & 0.00014 & 1 \\
2437 & 2459055.48789 & 0.00016 & 0.00016 & 2 \\
2439 & 2459058.45285 & 0.00027 & 0.00027 & 2 \\
2464 & 2459095.50861 & 0.00017 & 0.00017 & 1 \\
2466 & 2459098.47300 & 0.00022 & 0.00022 & 1 \\
2468 & 2459101.43743 & 0.00031 & 0.00031 & 1 \\
2499 & 2459147.38708 & 0.00016 & 0.00016 & 1 \\
2501 & 2459150.35166 & 0.00020 & 0.00020 & 1 \\
\hline
\end{tabular}
\tablefoot{$N_{\rm {lc}}$ is the number of light curves used.}
\end{table}

The transits observed by us were found to occur $\sim$2 minutes earlier than the ephemeris of \citet{2016AcA....66...55M} predicts. The refined $P_{\rm{orb}}$ was found to be shorter by $\sim$90~ms, which is a difference at a $5.8\sigma$ level. To investigate the cause of this discrepancy, we verified timestamps of all data used in previous transit studies, especially those at early epochs from \citet{2011ApJ...741..114M}. We found no mistake in the conversion to $\rm{BJD_{TDB}}$.

The timing residuals were analysed with the analysis of variance algorithm \citep[AoV,][]{1996ApJ...460L.107S} to search for the TTV signal reported by \citet{2018MNRAS.480..291S}. We employed a series of 3 Szeg{\"o} orthogonal polynomials that are equivalent to the usage of 1 harmonic, resulting in a Generalised Lomb-Scargle periodogram. The timing uncertainties were used to calculate weights. Levels of false alarm probability (FAP) were estimated with the bootstrap method based on $10^5$ trials. The periodogram is plotted in the lower panel of Fig.~\ref{fig:ttres}. Although there is no statistically significant signal at the frequency proposed by \citet{2018MNRAS.480..291S}, a low-frequency signal was found with FAP below 1\%. Its characteristics suggest that it could indeed be a periodic signal with a cycle duration above $\sim 10^4$~d or a quadratic trend produced by orbital period shortening.

\begin{figure}
   \centering
	\includegraphics[width=1\columnwidth]{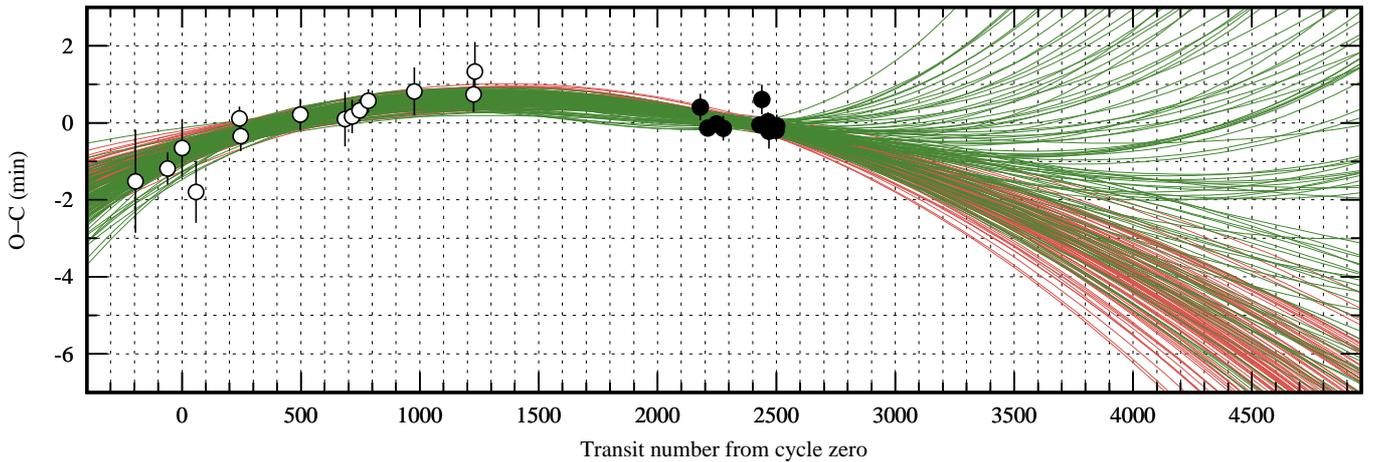}
    \caption{Timing residuals against the refined linear ephemeris. Data points are coded as in Fig.~\ref{fig:ttres}, and the data from \citet{2018MNRAS.480..291S} are skipped for clarity. The green and red lines show the periodic and quadratic-trend scenarios, respectively. Each bunch consists of 100 lines randomly picked from the Markov chains. Both scenarios remain statistically undistinguished in the time covered by current observations. The models are extrapolated to the next ten years to illustrate the broad spectrum of possible solutions.} 
    \label{fig:ttv}
\end{figure}

In the periodic scenario, the mid-transit times were represented by an ephemeris in the form  
\begin{equation}
  T_{\rm{mid}}= T_0 + P_{\rm{b}} E + A_{\rm TTV} \sin{\left[ 2\pi (E-\phi_{\rm TTV})/P_{\rm TTV} \right]},
\end{equation}
where $A_{\rm TTV}$, $P_{\rm TTV}$, and $\phi_{\rm TTV}$ are the amplitude of the TTV signal, its periodicity, and phase, respectively. The current timing data were found to impose weak constraints on those parameters, as shown in Fig~\ref{fig:ttv}. The MCMC-based method yields 68.4\% values of $P_{\rm TTV}$ in a range between 18000 and 40000~d around a median value of 31000~d. This finding is consistent with the results of the periodogram analysis. The best-fitting model has $\chi^2 = 9.8$ with 21 degrees of freedom.

In the quadratic-trend scenario, the transit ephemeris is given in the form
\begin{equation}
  T_{\rm{mid}}= T_0 + P_{\rm{b}} E + \frac{1}{2} \frac{{\rm d} P_{\rm{b}}}{{\rm d} E} E^2,
\end{equation}
where $\frac{{\rm d} P_{\rm{b}}}{{\rm d} E}$ is the change in the orbital period between succeeding transits. The best-fitting solution, which was found with the MCMC algorithm, yields $\frac{{\rm d} P_{\rm{b}}}{{\rm d} E} = (-9.6 \pm 2.2) \times 10^{-10}$ days per orbit, which translates into $\dot{P_{\rm{b}}} = \frac{1}{P_{\rm{b}}} \frac{{\rm d} P_{\rm{b}}}{{\rm d} E} = -20.4 \pm 4.7$ ms~yr$^{-1}$. The value of $\chi^2 $ was found to be equal to 14.7 with 23 degrees of freedom.

The Bayesian Information Criterion (BIC) was employed to assess the preferred model. Both the periodic and quadratic models were found to be favoured over the constant-period scenario with $\Delta {\rm BIC}$ equal to 13.3 and 14.9, respectively. However, the statistic does not make it possible to distinguish between the two nonlinear models for which $\Delta {\rm BIC}$ is just 1.6. This finding is illustrated in Fig.~\ref{fig:ttv}.

\section{Discussion}\label{Sect:Discussion}

The periodic transit timing signal that was reported by \citet{2018MNRAS.480..291S} would produce variations with a range of about 4 minutes. Our follow-up photometry provides transit timing precision at a level of about 20 s; thus, our data would easily detect the postulated signal if it existed. We notice that its amplitude is of the same magnitude as the spread of timing data in \citet{2018MNRAS.480..291S}. We attempted to reproduce the result of \citet{2018MNRAS.480..291S} by applying the methodology from Sect.~\ref{SSect:Results.Timing} to a subsample of the transit times used in the original study. A periodogram analysis revealed that the claimed 100-day periodicity is not the unique solution: a handful of similar power peaks at higher frequencies were identified. However, all of them, including the 100-day periodicity, were found to have disqualifying FAPs at a level of 50\%, which is much higher than the value of 0.18\% reported in the original study.

Although our transit timing analysis revealed no periodic perturbations of the orbital period for TrES-5~b, a sign of a departure from the linear ephemeris was detected. If we assume that this is a manifestation of orbital shortening due to tidal dissipation, the decay rate represented by $\frac{{\rm d} P_{\rm b}}{{\rm d} E}$ could be used to infer the value of the modified tidal quality parameter of the host star $Q'_{\star}$ \citep{1966Icar....5..375G}. This quantifies the effectiveness of tidal dissipation in the stellar interior. Using the constant-phase-lag model of \citet{1966Icar....5..375G}, in which
\begin{equation}
 Q'_{\star} = - \frac{27}{2}\pi \left( \frac{M_{\rm b}}{M_{\star}}\right) {\left( \frac{a}{R_{\star}} \right)}^{-5} {\left( \frac{{\rm d} P_{\rm b}}{{\rm d} E} \right)}^{-1} P_{\rm b} \, , \;
\end{equation}
we obtained $Q'_{\star} \approx 1.4 \times 10^4$. This result is smaller by several orders of magnitude than the values for typical dwarfs \citep[e.g.,][]{2017A&A...602A.107B} even if nonlinear dissipation processes are taken into account \citep{2016ApJ...816...18E}. Such a sharp decline would require invoking a novel mechanism that boosts the tidal dissipation in the host star. 

The apparent orbit shortening could be caused by an acceleration of the systemic barycentre along the line of sight. This manifestation of the Doppler effect (also known as the R{\o}mer effect, see \citet{2020ApJ...893L..29B} for a comprehensive discussion) yields a relation of the radial velocity (RV) acceleration $\dot{v}_{\rm RV}$ and $\dot{P}_{\rm b}$ with the formula
\begin{equation}
 \dot{v}_{\rm RV} = \frac{\dot{P}_{\rm b}}{P_{\rm b}} c \, , \;
\end{equation}
where $c$ is the speed of light. We obtained $\dot{v}_{\rm RV} = -0.13 \pm 0.03$ m~s$^{-1}$~day$^{-1}$. The only publicly available RV dataset come from \citet{2011ApJ...741..114M}. Eight measurements were performed between September 2010 and April 2011 with the 60 inch Tillinghast Reflector at Fred L.\ Whipple Observatory and the Tillinghast Reflector Echelle Spectrograph (TRES). Using the Systemic software \citep{2009PASP..121.1016M}, we determined the value of the barycentric acceleration from RV observations $\dot{\gamma} = -0.12 \pm 0.20$ m~s$^{-1}$~day$^{-1}$. The best-fitting solution was found with the Levenberg-Marquardt algorithm, and parameter uncertainties were determined with the bootstrap method with $10^5$ trials as median absolute deviations. In this model, the transit timing dataset was used to constrain the planet's orbital period, and the orbit was assumed to be circular. A short circularisation time scale justifies this assumption. Using Equation (25) of \citet{1966Icar....5..375G} and taking a typical value of the planetary tidal quality parameter equal to $10^6$, this scale was found to be of the order of $10^7$ yr, while the age of the system is specified as $(7.4 \pm 1.9) \times 10^9$ yr \citep{2011ApJ...741..114M}. Although the value of $\dot{\gamma}$ formally agrees with $\dot{v}_{\rm RV}$ well within $1\sigma$, its low precision calls for improvement, which can be achieved through long-term Doppler monitoring of the system.

Assuming that the orbital eccentricity $e_{\rm b}$ is non-zero, one could attempt to explain the periodic variations of $P_{\rm b}$ with a precession of apsides. Our joint RV-timing model yields the precession rate $\frac{{\rm d} \omega}{{\rm d}t} = 21.9 \pm 4.7$ degrees per year and $e_{\rm b} = 0.0016 \pm 0.0006$. The obtained value of $e_{\rm b}$ remains below detection thresholds in the current RV dataset. As the circularisation timescale is much shorter than the system's age, a non-circular orbit would require an exciting and sustaining mechanism that operated in the system. \citet{2009ApJ...698.1778R} showed that for very hot Jupiters, the planetary interior is a dominant component of the apsidal precession. The rate of this precession is related to the planetary Love number $k_{\rm 2p}$ which depends on the internal density distribution. Using Equation (14) of \citet{2009ApJ...698.1778R}, we obtained $k_{\rm 2p} \approx 8$ that is an unphysical result. This finding eliminates apsidal precession as a potential explanation for the departure from the linear transit ephemeris.
 
The Doppler acceleration of the system's barycentre would reflect its orbital motion induced by a massive, wide-orbiting companion. Thus the value of $\dot{P}_{\rm b}$ or $\dot{v}_{\rm RV}$ would be expected to be the subject of variability represented by the periodic scenario in a long timescale. Such a configuration would not be unusual. As discussed by \citet{2020ApJ...893L..29B}, wide-orbiting companions are expected to be detected in at least 31\% of systems with hot Jupiters. The current data allow us to estimate a minimum mass of the companion $M_{\rm c}$. Using Equation (8) from \citet{2020ApJ...893L..29B}, we derived $M_{\rm c} \approx 18 M_{\rm Jup}$. This result shows that the TrES-5 system might be orbited by a brown dwarf or a low mass star. \citet{2018MNRAS.480..291S} used speckle interferometry to eliminate stellar companions of up to 1 magnitude fainter than the host star for orbits between 72 and 1080 AU. While this result is sensitive to dwarfs down to early K spectral types, the regime of fainter stars and brown dwarfs remains unexplored.

To date, there are only two other hot Jupiters for which long-term changes in orbital periods have been detected: WASP-12~b \citep{2016A&A...588L...6M} due to tidal orbital decay \citep{2020ApJ...888L...5Y} and WASP-4~b \citep{2019AJ....157..217B} most likely due to line-of-sight acceleration \citep{2020ApJ...893L..29B}. The signs of the accelerations are expected to be equally distributed. Thus, as the time baseline for precise transit timing observations is continuously widened, hot Jupiters with positive period derivatives are predicted to also be identified.

\section{Conclusions}\label{Sect:Conclusions}

System parameters received from transit light curve modelling were found to agree with the results of previous studies. As the analysis of transit timing residuals against the linear ephemeris revealed no statistically significant signals with periods shorter than $10^4$~d, postulating the existence of a nearby perturbing planet appears to be premature.

Transit timing does, however, show a sign of a long-term trend that could be attributed to the apparent shortening of the orbital period of TrES-5~b. Although additional timing observations are required to confirm and strengthen this finding, we identify both orbital decay due to tides and apsidal precession as unlikely mechanisms which could produce the effect of the observed magnitude. Instead, we find the line-of-sight acceleration as a possible explanation for the period change. We note that precise Doppler monitoring and high angular resolution imaging, as well as precise transit timing observations, could help to explore this scenario. 

\begin{acknowledgements}
We thank Dr Georgi Mandushev for double-checking of time stamps used in the follow-up photometric time series. GM acknowledges the financial support from the National Science Centre, Poland through grant no. 2016/23/B/ST9/00579. MF acknowledges financial support from grant PID2019-109522GB-C5X/AEI/10.13039/501100011033 of the Spanish Ministry of Science and Innovation (MICINN). MF, FA, and JLR acknowledge financial support from the State Agency for Research of the Spanish MCIU through the \textit{Center of Excellence Severo Ochoa} award to the Instituto de Astrof\'{\i}sica de Andaluc\'{\i}a (SEV-2017-0709). This research has made use of the SIMBAD database and the VizieR catalogue access tool, operated at CDS, Strasbourg, France, and NASA's Astrophysics Data System Bibliographic Services.
\end{acknowledgements}

\bibliographystyle{aa} 
\bibliography{tr5_arxiv} 

\end{document}